\documentclass[a4paper,fleqn]{cas-sc}

\usepackage[authoryear]{natbib}
\usepackage{capt-of}

\def\tsc#1{\csdef{#1}{\textsc{\lowercase{#1}}\xspace}}
\tsc{WGM}
\tsc{QE}

\begin{document}
\let\WriteBookmarks\relax
\def\floatpagepagefraction{1}
\def\textpagefraction{.001}

\shorttitle{Hybrid QKD-PQC Network Emulation through Automated and Scalable Cloud-Native Orchestration}    

\shortauthors{}  

\title [mode = title]{Hybrid QKD-PQC Network Emulation through Automated and Scalable Cloud-Native Orchestration}  



%




\author[1]{Iván Melijosa}[orcid=0009-0007-2413-7529]
\cormark[1]
\ead{imelijos@pa.uc3m.es}

\author[1]{Javier Pérez}[orcid=0009-0008-3531-7696]
\ead{javierpe@pa.uc3m.es}

\author[1]{Borja Nogales}[orcid=0000-0002-5508-5414]
\ead{bdorado@pa.uc3m.es}

\author[1]{Iván Vidal}[orcid=0000-0001-7381-971X]
\ead{ividal@it.uc3m.es}

\author[1]{Francisco Valera}[orcid=0000-0001-5056-0573]
\ead{francisco.valera@uc3m.es}

\cortext[1]{Corresponding author}

\affiliation[1]{organization={Universidad Carlos III de Madrid},
            addressline={Av. de la Universidad, 30}, 
            city={Leganés},
            postcode={28911}, 
            state={Madrid},
            country={Spain}}







\begin{abstract}
The ongoing transition toward quantum-safe networking has motivated the development of hybrid network architectures integrating Quantum Key Distribution (QKD) and Post-Quantum Cryptography (PQC). However, the experimental evaluation of hybrid QKD-PQC network architectures remains constrained by the high cost and limited accessibility of quantum hardware, as well as by the limited support for hybrid QKD-PQC networks in existing emulation platforms. \textit{Quditto} is an open-source emulation platform originally designed for QKD networks that enables cost-effective and reproducible experimentation without requiring dedicated physical quantum infrastructure. Building on this foundation, this work presents \textit{Quditto} as a hybrid QKD-PQC network emulation platform featuring automated and scalable cloud-native orchestration. The proposed platform introduces four principal contributions: a cloud-native orchestrator enabling fully automated infrastructure deployment across cloud and multi-cluster environments; an optimized provisioning workflow enabling large-scale quantum-safe network emulation; native integration of post-quantum nodes enabling unified emulation of hybrid QKD-PQC networks; and a secure key management module providing persistent and access-controlled storage of cryptographic material. Experimental validation demonstrates sublinear orchestration-time scaling with network size and successful end-to-end hybrid QKD-PQC key establishment on a representative spine-leaf deployment, thereby enabling the systematic evaluation of quantum-safe networking mechanisms in large-scale heterogeneous network environments.
\end{abstract}


\begin{highlights}
    \item Cloud-native orchestration automating the scalable deployment of hybrid quantum-safe network emulation.
    \item Native support for hybrid QKD-PQC networking through standardized ETSI interfaces.
    \item Automated provisioning of virtual infrastructure across cloud virtual machines and Kubernetes clusters.
    \item \textit{HashiCorp Vault} ensures persistent, access-controlled storage of cryptographic material.

\end{highlights}



\maketitle

\section{Introduction}\label{sec:introduction}
The emergence of quantum computing represents a paradigm shift in cryptographic security, introducing unprecedented computational capabilities that challenge the foundations of existing information protection systems. In this context, quantum cryptography emerged from the pioneering work of Stephen Wiesner~\citep{wiesner} and gained prominence with the introduction of the BB84 protocol~\citep{bb84}, the first formally proposed Quantum Key Distribution (QKD) protocol. QKD leverages the principles of quantum mechanics to enable information-theoretically secure key distribution, ensuring that any eavesdropping attempt on the quantum channel inevitably introduces detectable disturbances~\citep{shor2000}. In contrast to classical cryptography, whose security relies on computational hardness assumptions, QKD derives its security guarantees from the laws of physics, rendering it inherently resilient to advances in both classical and quantum computing.

Paradoxically, the quantum mechanical principles that enable unconditionally secure key distribution also threaten the security of conventional cryptographic systems. Shor’s algorithm demonstrated that sufficiently powerful quantum computers could efficiently break widely deployed public-key cryptosystems, including RSA and Elliptic-Curve Cryptography (ECC)~\citep{shor1994}. Moreover, physical QKD devices remain difficult to integrate into existing communication infrastructures and are prohibitively expensive, restricting access to quantum-secured communications to a limited number of actors~\citep{pirandola2020}. These considerations have catalyzed the development of Post-Quantum Cryptography (PQC)~\citep{bernstein2009}, a field dedicated to designing cryptographic primitives resistant to classical and quantum attacks while maintaining compatibility with existing network infrastructures. The complementary properties of QKD and PQC have motivated research into hybrid QKD-PQC cryptographic architectures, which combine the information-theoretic security guarantees of QKD with the deployment flexibility of PQC to achieve quantum-safe communication across heterogeneous networks in which only a subset of network nodes is equipped with QKD capabilities.

In this technological context, the deployment of quantum cryptographic networks remains largely inaccessible to most researchers, limiting opportunities for cost-effective and reproducible large-scale experimentation. To address these challenges, \textit{Quditto}~\citep{quditto} was introduced as a Quantum Key Distribution Network (QKDN) emulation platform providing an accessible and controlled environment for studying QKD deployments without requiring dedicated physical quantum infrastructure. The platform can integrate high-fidelity simulation engines capable of reproducing the behavior of quantum devices, enabling experimentation under realistic operating conditions. In addition, \textit{Quditto} supports the seamless integration and interoperation of real-world deployments and emulated network environments within a unified experimental infrastructure distributed across heterogeneous physical and virtual computing systems \citep{testbed}.

However, the original \textit{Quditto} platform did not support the emulation of hybrid QKD-PQC network architectures, hindering the experimental evaluation of emerging quantum-safe networking solutions. Moreover, the original design assumed the availability of pre-existing, network-accessible devices and required manual infrastructure provisioning, preventing the platform from leveraging modern cloud-native technologies~\citep{deng2023}, which enable the automated deployment, configuration, scaling, and management of containerized applications across distributed computing infrastructures through declarative orchestration frameworks, thereby providing lightweight, portable execution environments and distributed resource management. Furthermore, the original deployment workflow relied on sequential runtime software provisioning and node initialization, introducing orchestration bottlenecks that constrained the practical scalability of the emulation environment. Building upon these aspects, the present work presents \textit{Quditto} as a hybrid-capable, autonomous, and scalable emulation platform with the following contributions:
\begin{itemize}
    \item Native integration of post-quantum nodes implementing IKEv2-based key establishment with ML-KEM and ML-DSA algorithms, enabling unified emulation of hybrid QKD-PQC network topologies.
    \item A cloud-native orchestrator enabling the fully automated deployment of the underlying virtual infrastructure across cloud-hosted virtual machines and multi-cluster Kubernetes environments.
    \item A performance-optimized deployment workflow based on prebuilt container images and parallelized provisioning and initialization operations, enabling the deployment of large-scale emulated quantum-safe networks comprising up to 200 nodes in approximately 6 minutes.
    \item A secure cryptographic material management module providing persistent, access-controlled storage based on the secrets management service \textit{HashiCorp Vault}~\citep{vault}.
\end{itemize}

The proposed platform enables the research community to systematically evaluate quantum-safe cryptographic mechanisms in hybrid QKD-PQC network environments at scale and under realistic, reproducible deployment conditions, helping bridge the gap between theoretical architectural design and experimental validation. Concretely, the platform enables experiments including the evaluation of QKD protocols under configurable channel conditions and eavesdropping models, compliance testing of key management interfaces against standardized API specifications, functional validation of hybrid QKD-PQC key establishment workflows across heterogeneous node configurations, and performance benchmarking of quantum-safe cryptographic mechanisms under controlled network load. These experiments can be conducted on large-scale emulated quantum-safe networks deployable on standard virtual infrastructure within practical deployment times, without requiring dedicated quantum hardware or physical testbeds.

The present work is structured as follows. Section~\ref{sec:StateOfTheArt} reviews three research areas: hybrid QKD-PQC network architectures, existing emulation platforms, and orchestration frameworks. Sections~\ref{sec:Design} and~\ref{sec:Implementation} describe the design and implementation of \textit{Quditto}. Section~\ref{sec:Validation} presents the experimental evaluation of the platform, covering orchestration performance and functional validation. Finally, Section~\ref{sec:Conclusions} concludes the paper and outlines future research directions.

\section{State of the Art}\label{sec:StateOfTheArt}
This section reviews the state of the art across three areas: hybrid QKD-PQC network architectures, quantum-safe network emulation platforms, and orchestration frameworks for automated and scalable deployment. The limitations identified in the reviewed literature motivate the design objectives of the emulation platform presented in this work.

\subsection{Hybrid QKD-PQC Network Architectures} \label{sec:hybrid}
\begin{figure*}
    \centering 
    \includegraphics[page=1,width=0.8\linewidth]{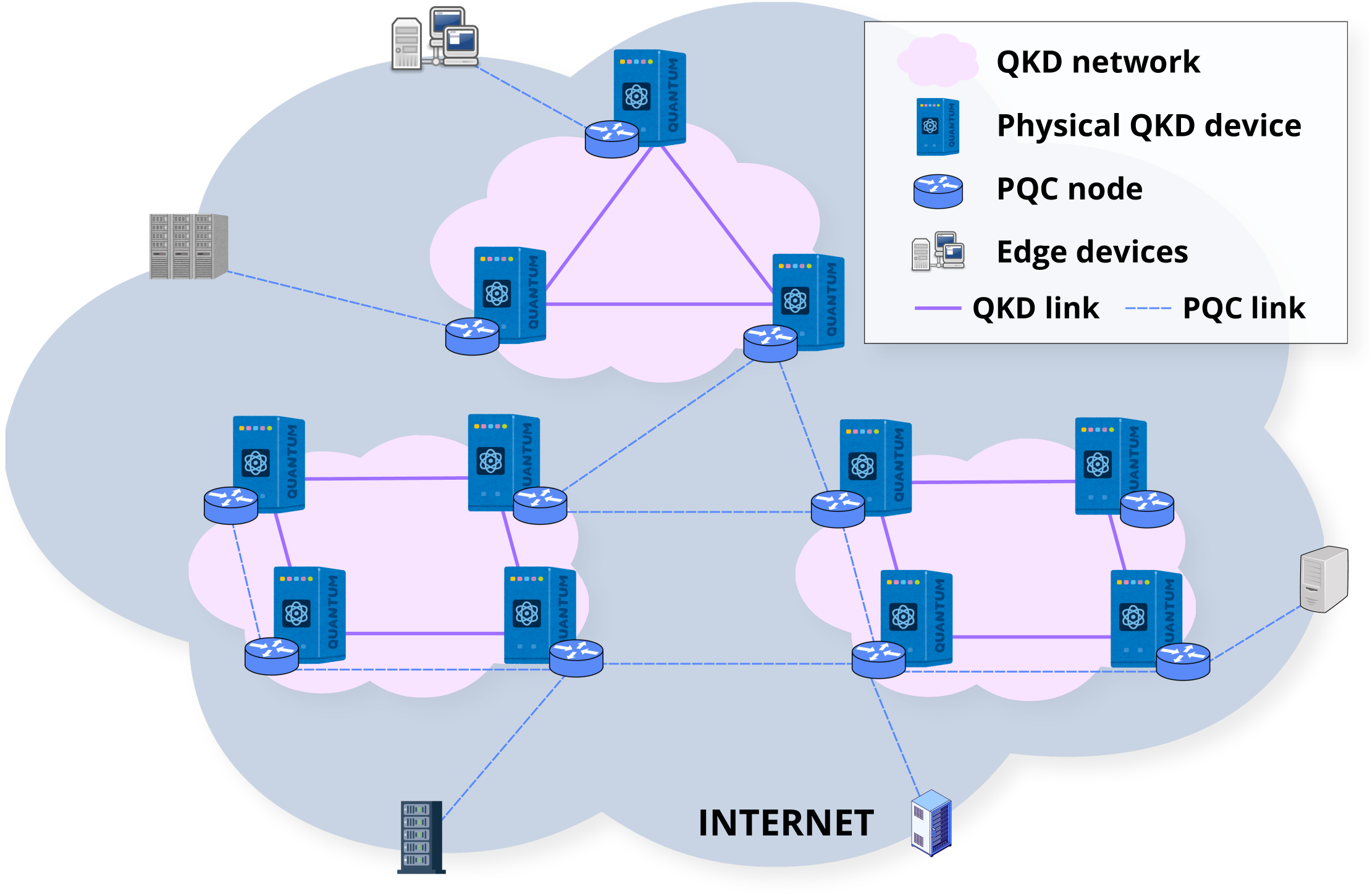}
    \caption{Hybrid QKD-PQC network architecture.}
    \label{fig:hybrid_network}
\end{figure*}

Hybrid network architectures integrating QKD and PQC have been proposed to address the inherent limitations of each technology when considered in isolation~\citep{campagna2015}. QKD offers information-theoretic security guarantees; however, its practical deployment faces significant scalability challenges and typically requires dedicated optical infrastructure~\citep{mehic2020}. In contrast, PQC supports large-scale deployment over existing communication networks, although its security relies on the assumed computational hardness of specific mathematical problems rather than physical principles, rendering it potentially vulnerable to future algorithmic advances~\citep{bernstein2017}. The complementary security and deployment properties of QKD and PQC have motivated the development of hybrid QKD-PQC network architectures that integrate both technologies to achieve quantum-safe communication while preserving deployment flexibility. As illustrated in Fig.~\ref{fig:hybrid_network}, these architectures typically comprise multiple QKDN segments integrated within a broader classical communication infrastructure and interconnected through PQC nodes. QKD is generally reserved for high-value core network segments requiring unconditional security guarantees and relying on trusted relay nodes, which decrypt and re-encrypt cryptographic keys at each hop and therefore must operate within physically secure environments, whereas PQC nodes extend quantum-safe cryptographic protection across inter-segment connections and toward heterogeneous edge devices over existing communication infrastructure~\citep{spooren2025, comin2025}, particularly where QKD deployment is either unavailable or economically impractical.

The importance of hybrid QKD-PQC cryptographic architectures has been further reinforced by recent standardization progress. The National Institute of Standards and Technology (NIST) published its first PQC standards, FIPS 203 and FIPS 204, defining the Module-Lattice-Based Key Encapsulation Mechanism (ML-KEM)~\citep{fips203} and the Module-Lattice-Based Digital Signature Algorithm (ML-DSA)~\citep{fips204} as the primary mechanisms for post-quantum key establishment and digital signature, respectively. In parallel, regulatory bodies have established migration roadmaps to guide the transition to PQC to ensure long-term cryptographic security~\citep{pqc-transition}, catalyzing coordinated efforts to incorporate post-quantum cryptographic mechanisms into existing communication protocols. In this context, the Internet Engineering Task Force (IETF) has proposed extensions to the Internet Key Exchange version 2 (IKEv2) protocol~\citep{rfc7296} supporting hybrid and multiple key-establishment mechanisms~\citep{rfc9370}, while additional post-quantum extensions remain under active development within IETF working groups~\citep{ietf-pquip-charter}. The European Telecommunications Standards Institute (ETSI) has standardized architectural frameworks for QKDNs through specifications such as ETSI GS QKD 004~\citep{etsi-qkd004}, which defines the internal interface between a Key Management Entity (KME) and its associated QKD module, and ETSI GS QKD 014~\citep{etsi-qkd014}, which specifies the application-facing REST interface through which Secure Application Entities (SAEs) request key material. Within this architecture, KMEs coordinate and distribute quantum-generated key material, while SAEs consume these keys at the application layer. Furthermore, ETSI has published specifications addressing the integration of hybrid quantum-safe key establishment mechanisms within quantum-safe communication infrastructures~\citep{etsi-ts103744,etsi-tr103966}, enabling recent research on adaptive hybrid frameworks for heterogeneous QKD-PQC environments~\citep{adaptive-framework}.

Despite these advances, several challenges remain unresolved. At the architectural level, the reliance on trusted relay nodes, stemming from the limited transmission range of current QKD systems and the absence of practical quantum repeaters, and the complexity of managing unified keying material across heterogeneous network segments represent only some of the major engineering challenges facing hybrid quantum-safe networks. More critically, the surveyed proposals above predominantly address architectural design and limited-scope proof-of-concept validation, leaving open the challenge of systematic experimental evaluation of hybrid QKD-PQC networks at scale. In practice, validation is typically performed through small-scale physical testbeds, which are costly, difficult to reproduce, and largely inaccessible to the broader research community. This highlights the need for dedicated emulation infrastructures capable of reproducing hybrid QKD-PQC network behavior under realistic operating conditions.

\subsection{Emulation Platforms for Quantum-Safe Networks} \label{subsec:emulation}

The high cost and complexity of deploying physical QKD infrastructure have motivated the development of a broad range of simulation and emulation tools for QKDNs, operating at different levels of abstraction. While simulation approaches rely on abstract models to study network behavior, emulation approaches execute network components in controlled environments that reproduce real deployment conditions more closely. At the device level, NetSquid~\citep{netsquid} provides a discrete-event simulation engine capable of modeling individual quantum hardware components
with high physical fidelity, including realistic noise and decoherence effects. While this level of detail makes NetSquid well-suited for prototyping new QKD protocols and studying device behavior under realistic physical conditions, its scope remains limited to individual device interactions and channel physics, and it does not natively support distributed network topologies or standardized key-delivery interfaces. At the network layer, platforms such as SimQN~\citep{simqn}, SeQUeNCe~\citep{sequence}, and QuNetSim~\citep{qunetsim} abstract away
low-level quantum dynamics to enable the study of routing strategies, key management policies, and protocol behavior at scale. Similarly, QKDNetSim~\citep{qkdnetsim} and its enhanced successor QKDNetSim+~\citep{qkdnetsimplus} extend the widely used NS-3 network simulator~\citep{ns3} with a dedicated QKD module implementing ETSI-compliant key management functionality, enabling network-level experiments with configurable QKD topologies. More recently, work by Mehic
et al.~\citep{mehic2025emulation} has pushed this line of research toward emulation, presenting an extended QKDNetSim-based ecosystem for the virtual deployment of a national QKDN, bridging the gap between simulation and real-world operation. Collectively, these platforms offer complementary perspectives on QKDN behavior, yet all are fundamentally limited for many different reasons: none of them provide mechanisms for modeling post-quantum cryptographic components or the interactions that arise in hybrid QKD-PQC network configurations; none are capable of creating distributed nodes on a real network, only working with local operations; and more importantly, they are not capable of working collectively with real quantum or post-quantum hardware.

On the post-quantum side, simulation efforts have focused primarily on protocol-level evaluation and performance benchmarking of PQC algorithms in isolated settings rather than on the emulation of distributed PQC network infrastructures~\citep{pqcan}. Physical testbeds such as the Berlin OpenQKD deployment~\citep{berlinopenqkd} have demonstrated the 
feasibility of integrating QKD hardware with PQC-based key management systems in real-world environments. However, these deployments are costly, difficult to reproduce, and inherently limited in scale, making 
them unsuitable as general-purpose experimental platforms for the systematic evaluation of hybrid QKD-PQC network behavior. Consequently, none of the existing simulation, emulation, or physical testbed approaches natively support the integration of post-quantum components with real quantum hardware at scale, preventing researchers from conducting controlled and reproducible experimentation on hybrid network behavior. 

Collectively, these limitations constitute the primary gap addressed by this work: the absence of a scalable emulation platform for the systematic evaluation of architectural designs, key management strategies, and protocol interactions in hybrid QKD-PQC network environments.

\subsection{Orchestration Frameworks for Emulation Platforms}

Network deployment automation has evolved substantially with the widespread adoption of containerization technologies and Infrastructure as Code practices. By packaging applications together with their software dependencies into lightweight, portable execution environments, containerization has simplified the deployment of distributed systems across heterogeneous computing infrastructures. As deployments increased in complexity and scale, container orchestration platforms, particularly Kubernetes~\citep{kubernetes}, have become the default standard for automating the deployment and management of containerized applications, providing
native support for automated scaling, declarative configuration whereby application requirements are expressed as target system states rather than procedural instructions, and resource management across heterogeneous compute environments. Complementary infrastructure provisioning tools, such as Terraform~\citep{terraform}, address the provisioning layer of the deployment lifecycle by enabling the reproducible creation of virtual infrastructure across cloud environments through Virtual Infrastructure Managers (VIMs) such as OpenStack~\citep{openstack}. Configuration management and software deployment can subsequently be automated using dedicated tools, such as Ansible~\citep{ansible}, which automates software installation and service initialization across target nodes, or through Kubernetes-native mechanisms. Together, these technologies have established a well-integrated automation stack that supports reproducible infrastructure deployment across diverse research and production environments.

Extensions to this automation stack have emerged to address specific requirements in heterogeneous deployments. KubeVirt~\citep{kubevirt} bridges the boundary between virtualization and containerization by allowing virtual machines, which emulate complete operating systems, to coexist and be managed alongside lightweight containerized applications within the same Kubernetes cluster. Both execution environments are orchestrated through the Kubernetes control plane, the centralized management component responsible for scheduling and coordinating workloads across the cluster, thereby providing a unified orchestration interface for mixed infrastructure environments. Similarly, cross-cluster networking solutions such as Submariner~\citep{submariner} extend Kubernetes networking across independent cluster boundaries, enabling direct pod-to-pod communication by providing pod-level IP reachability and service discovery across multi-cluster deployments.

Despite this breadth of tooling, existing orchestration frameworks exhibit critical limitations when applied to the specific requirements of quantum-safe network experimentation. Although solutions such as KubeVirt provide unified orchestration for virtual machines and containerized applications, they lack declarative mechanisms for describing heterogeneous network topologies spanning both execution environments within a single deployment configuration. Furthermore, as evidenced by the platforms surveyed in Section~\ref{subsec:emulation}, no existing framework provides integrated support for quantum channel simulation combined with post-quantum cryptographic protocol deployment within a unified experimental environment. Consequently, these gaps collectively constrain the ability to systematically design, deploy, and evaluate quantum-safe network architectures across heterogeneous infrastructure configurations without requiring access to dedicated quantum hardware or specialized infrastructure expertise, while nonetheless allowing the creation of mixed virtual-physical testbeds whenever such resources are available.

\section{Design} \label{sec:Design}
Building upon the foundational principles of the original platform, \textit{Quditto} preserves its core design objectives: distributed emulation capabilities, adherence to standardized key-delivery interfaces, flexibility to support realistic modeling of quantum networks across heterogeneous technologies and protocols, and accessibility to users regardless of programming background. Beyond these inherited principles, the proposed platform introduces two additional capabilities to address the aspects identified in Section \ref{sec:StateOfTheArt}. First, it provides native support for the definition and operation of hybrid network configurations integrating QKD and PQC components. Second, it enables
the automated and scalable deployment of the underlying virtual infrastructure on which network emulation environments are dynamically instantiated to reproduce diverse hybrid network scenarios. These capabilities remain largely absent from existing emulation platforms, which typically neither support the specification and execution of hybrid QKD-PQC network topologies nor provide automated mechanisms for provisioning and managing the virtual computing resources required for large-scale network emulation. Collectively, they establish a unified, scalable, and reproducible platform for emulating hybrid quantum-safe networks under realistic deployment conditions.

\begin{figure*}
    \centering 
    \includegraphics[page=1,width=\linewidth]{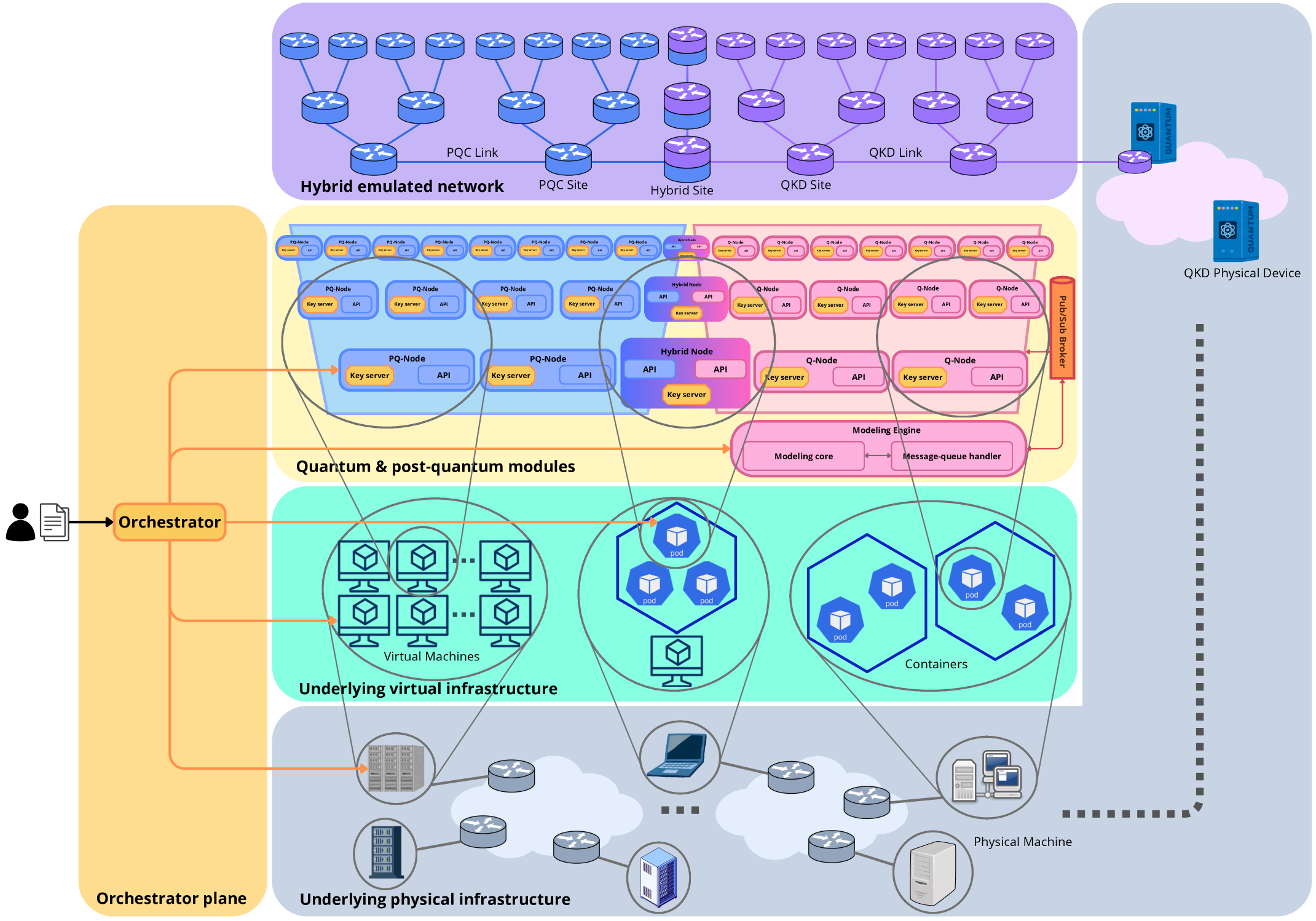}
    \caption{\textit{Quditto} architectural design.}
    \label{fig:design}
\end{figure*}

The proposed \textit{Quditto} platform is organized into four hierarchical layers coordinated by the orchestrator, as illustrated in Fig.~\ref{fig:design}. The orchestrator operates as a cross-cutting management plane that coordinates provisioning and deployment operations across all architectural layers from a user-provided configuration. The first layer comprises the \textit{underlying physical infrastructure}, including the physical machines and network equipment over which the platform is deployed and to which the orchestrator maintains connectivity. The second layer corresponds to the \textit{underlying virtual infrastructure}, where virtual machines and containerized environments are provisioned and managed by the orchestrator on top of the underlying physical infrastructure. The third layer encompasses the \textit{quantum and post-quantum modules}, comprising \textit{Quditto} nodes supporting QKD or PQC functionality and the \textit{Quditto} modeling engine, all instantiated as independent virtualized components within the underlying virtual infrastructure. Each node exposes a standardized key-delivery interface and incorporates a secure key management component for the protected storage of cryptographic material, whereas the modeling engine comprises a modeling core and a message-queue handler for quantum channel simulation. Finally, the fourth layer represents the \textit{hybrid emulated network}, an abstract topology of interconnected sites defining the logical network structure over which quantum-safe cryptographic operations are performed. The platform additionally supports transparent integration with physical QKD devices through a dedicated interface, bridging emulated and real-world quantum network deployments and enabling progressive migration toward physical quantum infrastructure. The following subsections provide a detailed description of each architectural layer of the platform.

\subsection{\textit{Quditto} Orchestrator}
The \textit{Quditto} orchestrator supports the automated deployment of both the underlying virtual infrastructure and the network emulation environment directly from a user-provided declarative configuration, without requiring pre-existing virtual resources. In this context, the \textit{underlying virtual infrastructure} comprises the virtual computing, storage, and networking resources required to host and execute \textit{Quditto}, instantiated as virtual machines in cloud environments or as container-based clusters in cloud-native deployments. On top of this infrastructure, the \textit{network emulation environment} comprises the virtual network components representing the communication elements of the emulated network.

At the architectural level, the orchestrator enforces a clear separation between infrastructure provisioning and network emulation orchestration, while enabling their unified management through a common interface. Once the virtual infrastructure has been provisioned, the orchestrator instantiates the distributed \textit{Quditto} platform, creating the network emulation environment corresponding to the user-defined network topology. This two-stage deployment process, comprising infrastructure and emulation environment instantiation, results in a fully operational distributed network deployment.

Beyond the initial specification of the infrastructure configuration and the target network topology, the orchestration workflow requires no further manual intervention. The orchestrator supports both fully automated end-to-end deployment and integration with user-managed infrastructure, allowing flexible operation across diverse execution environments while ensuring the accessibility and reproducibility of experimental results. To support large-scale network emulation, the orchestrator is designed to operate across multi-cluster environments, enabling distributed resource provisioning and management, as discussed in Section~\ref{sec:infrastructure}.

\subsection{Underlying Infrastructure} \label{sec:infrastructure}
The underlying infrastructure supporting \textit{Quditto} comprises two layers: a physical infrastructure and a virtual infrastructure built on top of it to host the platform components. The physical infrastructure constitutes the pre-existing physical substrate on which the platform is deployed and comprises computing and networking resources, including servers, networking equipment, and the connectivity required by the orchestrator. On top of this layer, the orchestrator instantiates and manages the underlying virtual infrastructure, which abstracts the available computing resources into virtualized deployment environments suitable for network emulation. The platform supports three deployment configurations within this virtual infrastructure: a virtual machine-based environment, where the platform is deployed across multiple virtual machines; a heterogeneous environment combining virtual machines with a container-based cluster; and a distributed multi-cluster container environment spanning multiple independent container clusters. This abstraction enables the same orchestration workflow to operate across different infrastructure configurations while providing a common execution environment for the upper architectural layers.

To overcome the scalability limitations observed in the original platform, \textit{Quditto} supports container-based cluster orchestration as a deployment configuration for large-scale network emulation environments. The original platform relied on a virtual machine-based infrastructure, which limited efficient deployment due to the resource overhead associated with provisioning and managing independent virtual machine instances. Containerization provides several advantages over traditional virtual machine-based deployments in the context of large-scale network emulation. First, containers have a significantly reduced resource footprint, since they share the host operating system kernel rather than requiring dedicated operating system instances, enabling higher deployment densities and allowing a single physical host to support a larger number of virtual instances. Second, containers provide faster instantiation times of software components, typically on the order of seconds rather than minutes, thus reducing the latency associated with network deployment operations. Third, containerization improves environment portability across heterogeneous infrastructure by encapsulating software dependencies within standardized container images that can be deployed consistently across different host systems. Finally, container orchestration frameworks provide mechanisms for declarative configuration, automated deployment, scaling, and distributed resource management, thereby simplifying the provisioning and operation of large-scale emulation environments. Consequently, the adoption of container-based clusters establishes the architectural foundation for the scalability evaluation presented in Section~\ref{sec:Validation}, enabling \textit{Quditto} to support large-scale hybrid QKD-PQC network emulation scenarios under realistic deployment conditions.

\subsection{Network Emulation Environment}
\begin{figure}
    \centering 
    \includegraphics[page=1,width=0.9\linewidth]{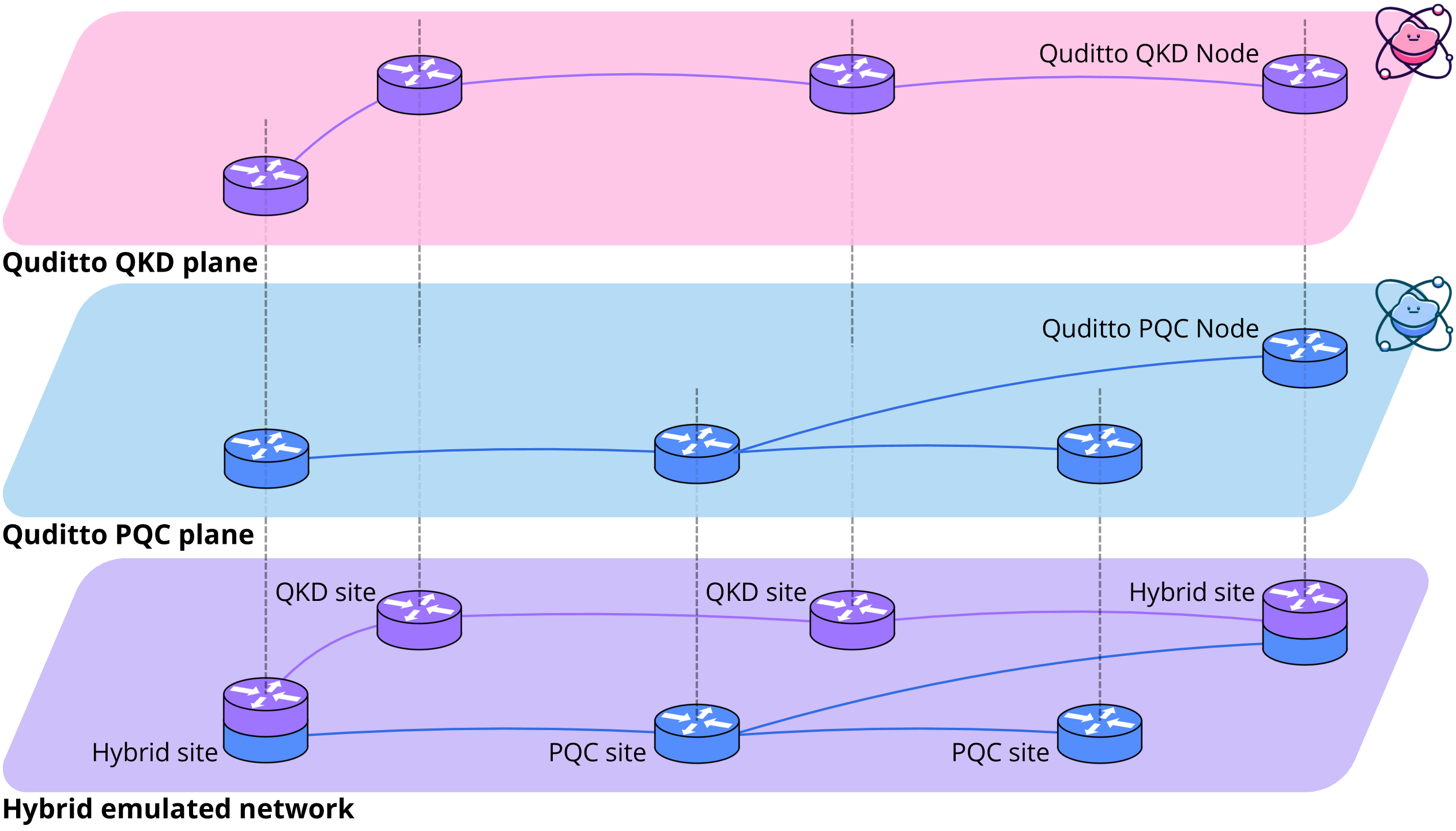}
    \caption{Architectural node-based overview of the \textit{Quditto} QKD-PQC hybrid network architecture.}
    \label{fig:node_design}
\end{figure}

The network emulation environment comprises the virtual components required to reproduce the communication elements of the emulated network, including \textit{Quditto} nodes and the \textit{Quditto} modeling engine. The \textit{Quditto} nodes are digital representations of physical network devices responsible for handling cryptographic material requests from client applications and external entities, including KMEs and SAEs, through a standardized key-delivery API. The \textit{Quditto} modeling engine is responsible for reproducing the behavior of quantum communication processes through high-fidelity simulations of quantum channels and QKD protocols, comprising a message-broker handler for asynchronously managing simulation tasks and results, and a modeling core for executing the underlying quantum communication models. Each \textit{Quditto} node operates either in QKD mode or in PQC mode, according to the cryptographic functionality it provides. Consequently, every node is associated with one of two distinct and logically separated cryptographic network planes, corresponding to QKD and PQC, respectively. In this context, a \textit{site} denotes a logical location that hosts one QKD node, one PQC node, or both simultaneously, in which case it is referred to as a hybrid site, whereas a \textit{node} refers to the individual cryptographic-plane component itself. Together, these planes define the logical topology of the hybrid emulated network comprising abstract QKD-only, PQC-only, and hybrid sites, as illustrated in Fig.~\ref{fig:node_design}.

Prior to deployment, the topology of each plane is independently defined, providing flexibility in configuring node placement, interconnections, and link characteristics. Once deployed, each plane operates autonomously; however, both planes can be instantiated simultaneously to enable a hybrid key establishment environment that preserves strict logical separation between the two cryptographic mechanisms within the overall network architecture. This separation ensures that each key establishment mechanism operates independently while still supporting coordinated key distribution workflows when required.

The \textit{Quditto} QKD plane comprises a virtualized network of \textit{Quditto} nodes capable of processing requests for quantum cryptographic key material. These nodes handle requests for the generation and storage of quantum cryptographic keys with neighboring nodes, as well as requests for the retrieval of previously generated keys, using unique key identifiers to ensure unambiguous association between the generated key material and the corresponding key generation session. Key material is generated by the \textit{Quditto} modeling engine through high-fidelity simulation of QKD protocols and quantum channels, reproducing the statistical properties and operational behavior of real-world quantum communication systems, including channel error rates and key generation dynamics. The generated key material is then asynchronously delivered to the nodes through a message-broker-based communication interface.

The \textit{Quditto} PQC plane comprises distributed \textit{Quditto} nodes emulating post-quantum network devices capable of establishing quantum-safe shared secrets through a pairwise, mutually authenticated key negotiation framework. Each key establishment session follows an initiator-responder model in which mutual authentication is achieved through a post-quantum digital signature scheme, assuming that the public key of each node is certified by a trusted Certificate Authority (CA), thereby establishing a PKI-based trust model. The key establishment mechanism is based on the IKEv2 protocol, conventionally used to negotiate Security Associations (SAs) and establish cryptographic keying material for IPsec communications. In the proposed architecture, IKEv2 is adapted as a standalone key establishment service, decoupling its cryptographic key derivation procedures from IPsec tunnel configuration and data-plane functions, and extended to incorporate post-quantum algorithms. As a result, IKEv2 operates as an on-demand mechanism for generating quantum-safe shared secrets while preserving its original security properties, including mutual authentication, key freshness, and resistance to replay attacks. IKEv2 was selected over alternative key establishment protocols, such as TLS 1.3~\citep{tls}, because its peer-to-peer key establishment model closely matches the ETSI QKD architecture, where KMEs establish secure associations independently of application traffic. Furthermore, IKEv2 provides a mature and extensible framework for post-quantum integration through ongoing IETF standardization efforts, facilitating interoperability while preserving compatibility with existing security architectures. Upon successful key establishment, the initiator generates a random key identifier and transmits it to the responder, enabling both endpoints to unambiguously associate the derived keying material with a common reference identifier. This mechanism mirrors the identifier-based key management model of the QKD plane, in which both endpoints independently possess identical cryptographic material bound to a common identifier. By adopting a consistent identifier scheme across both planes, the system enables a unified key management model in which client applications retrieve key material through the same API, regardless of the underlying cryptographic mechanism.

At runtime, client applications submit key requests to nodes in either plane via the standardized ETSI GS QKD 014 API. Each node internally routes the request to the appropriate cryptographic plane and handles it according to the configured key establishment mechanism. The platform natively supports multiple protocols and can be extended through custom implementations, enabling the integration of additional mechanisms while ensuring consistent handling across heterogeneous hybrid infrastructure.

To provide persistent and secure management of cryptographic material, each \textit{Quditto} node integrates a dedicated secure key management component responsible for its storage, protection, and retrieval. The original platform relied on in-memory storage for cryptographic material, which provided neither persistence across node restarts nor protection of key material at rest without mechanisms for access traceability. The proposed architecture addresses these limitations by introducing a secure key management layer that provides persistent, integrity-protected, and auditable management of cryptographic material throughout its entire lifecycle.

The secure key management component supports the storage and retrieval of cryptographic keys generated by both the QKD and PQC planes. In addition to cryptographic key material, it stores the SA state associated with post-quantum key establishment sessions, enabling efficient stateful rekeying. All stored material is protected at rest and is accessible exclusively through authenticated interfaces subject to fine-grained access control policies, ensuring that access is restricted to authorized entities. This unified storage layer abstracts the differences between the two cryptographic planes, exposing a consistent retrieval interface to client applications regardless of the underlying key generation mechanism. By enforcing fine-grained access control at the node level, the proposed design reduces the attack surface of the platform while ensuring that cryptographic material remains accessible only to authorized entities through auditable and traceable operations. These properties establish secure key management as a core security component of the platform, enabling consistent and secure end-to-end management of cryptographic material across heterogeneous cryptographic planes.

\section{Implementation} \label{sec:Implementation}
\textit{Quditto} is implemented as a Python-based open-source platform with dedicated packages for its three core components: the nodes, the modeling engine, and the orchestrator. The implementation details of these packages are described in the following subsections.

\subsection{\textit{Quditto} Node Package}
\begin{figure}
    \centering
    \includegraphics[page=1,width=\columnwidth]{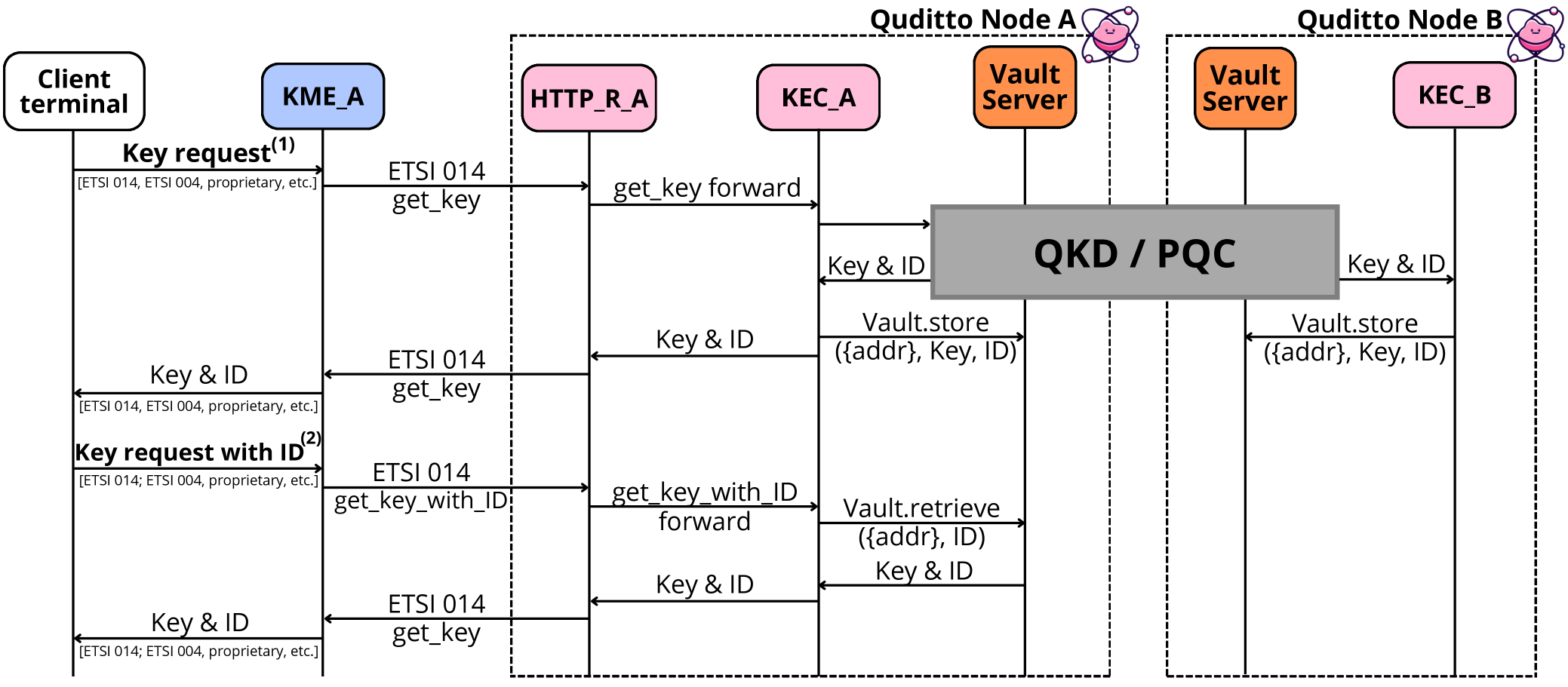}
    \caption{Key establishment sequence diagram for \textit{Quditto} nodes, illustrating the interaction between the Key Establishment Component (KEC), which can correspond to either a QKD or PQC component, and the HTTP receptor responsible for processing key establishment requests from KMEs.}
    \label{fig:key_exchange}
\end{figure}

The \textit{Quditto} node package implements two principal components: a quantum component for emulating quantum network nodes and a post-quantum component for emulating post-quantum network nodes. Additionally, each node runs a local \textit{HashiCorp Vault} instance for the secure and unified management of cryptographic material across both QKD and PQC planes. The complete key establishment workflow, including Vault-based key storage and retrieval, is illustrated in Fig.~\ref{fig:key_exchange}.

Upon a client-issued \texttt{get\_key} request (1), the KME forwards an ETSI GS QKD 014 \texttt{get\_key} request to the HTTP receptor of \textit{Quditto} Node A, which triggers the Key Establishment Component (KEC) to initiate either a QKD- or PQC-based key establishment procedure with the peer node. The resulting cryptographic key and its associated identifier are securely stored in the local Vault instance at both endpoints before being returned to the client through the standardized interface. For subsequent \texttt{get\_key\_with\_ID} requests (2), the HTTP receptor retrieves the corresponding key from the local Vault instance using the provided identifier. Both request types are accessible through multiple interfaces, including the standardized ETSI GS QKD 004 and 014, as well as proprietary interfaces.

The quantum component emulates a QKDN node through two concurrent processes. The first process is responsible for communication with the \textit{Quditto} modeling engine through a RabbitMQ-based~\citep{rabbitmq} message broker, which relays key generation requests and returns the resulting quantum cryptographic material. The second process hosts an HTTP server exposing the ETSI GS QKD 014-compliant API, enabling client applications to request cryptographic material using standardized primitives according to their role in the key establishment procedure: \texttt{get\_key} for the initiating endpoint and \texttt{get\_key\_with\_ID} for the responding endpoint, which retrieves the previously established key using its identifier. If a request containing a key identifier is received, the node searches for the corresponding entry in its local key store. If found, the associated key material is returned directly without interacting with the modeling engine. Keys are subject to a configurable time-to-live (TTL), with a default value of 10 minutes, consistent with the operation of commercial QKD devices.

The post-quantum component emulates a post-quantum network node. Compared with the quantum component, the communication architecture is simplified. Since no interaction with the modeling engine is required, the RabbitMQ-based message broker is omitted and replaced by a direct socket-based client-server communication model. As with the quantum component, the HTTP server exposing the ETSI GS QKD 014-compliant API handles requests for post-quantum cryptographic material. In the post-quantum case, the key size parameter is not applicable, since ML-KEM produces fixed-length shared secrets.

\begin{figure}
\centering    \includegraphics[width=0.619\linewidth]{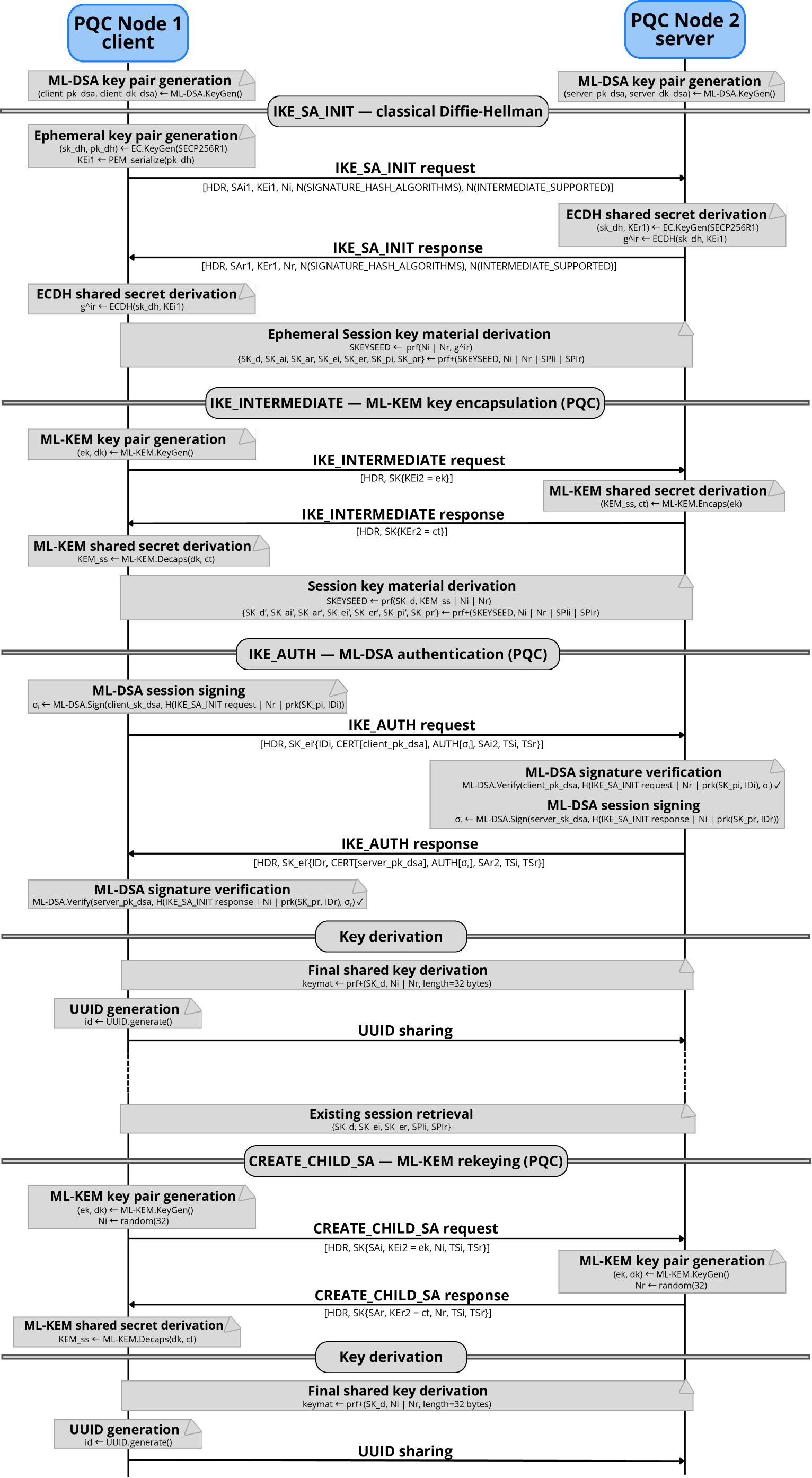}
    \caption{Post-quantum key establishment sequence diagram between \textit{Quditto} PQC nodes.}
    \label{fig:pqc_exchange}
\end{figure}

At the cryptographic level, post-quantum nodes implement key establishment using the IKEv2 protocol. IKEv2 packet construction and processing are implemented using the Scapy Python library~\citep{scapy}, which was selected for its flexibility in constructing, parsing, and extending protocol packets. Classical cryptographic operations, including ECC key generation, HKDF-based key derivation, hashing, and AES-GCM encryption, are implemented using the Python Cryptography library~\citep{cryp-lib}. Post-quantum cryptographic primitives are implemented using the reference implementations of ML-KEM~\citep{gp-mlkem} and ML-DSA~\citep{gp-mldsa}, with minor modifications to ensure compliance with FIPS 203 and FIPS 204, respectively. Building on this cryptographic foundation, the protocol-level implementation extends the standard IKEv2 protocol within the unified request-driven model through post-quantum mechanisms based on two IETF Internet-Drafts. The first defines hybrid key exchange by combining classical Diffie-Hellman with ML-KEM~\citep{ikev2-mlkem}, while the second integrates ML-DSA into the IKEv2 authentication framework~\citep{ikev2-mldsa}. The complete post-quantum key establishment sequence is illustrated in Fig.~\ref{fig:pqc_exchange}.

The initial key establishment between client and server proceeds through three sequential phases. During \textit{IKE\_SA\_INIT}, both parties perform an ephemeral Elliptic Curve Diffie-Hellman (ECDH) key exchange, deriving a shared secret used to generate the initial IKEv2 keying material. During \textit{IKE\_INTERMEDIATE}, ML-KEM key encapsulation is performed. The client provides its encapsulation key, and the server returns the resulting ciphertext. Both parties then derive the corresponding shared secret and use it to update the IKEv2 keying material, thereby injecting the post-quantum contribution into the IKEv2 key derivation session. During \textit{IKE\_AUTH}, mutual authentication is performed using ML-DSA. Each party signs the session transcript with its private key and verifies the signature received from its peer, thereby authenticating both endpoints and completing the key establishment procedure. Upon completion, the SAs are established to maintain the session state across subsequent key establishment operations, and the final shared key is derived from the accumulated IKEv2 keying material. The client then generates a UUID, which is transmitted to the server to establish a shared key identifier for the resulting cryptographic material. For subsequent key refresh operations, a single \textit{CREATE\_CHILD\_SA} exchange performs ML-KEM-based rekeying, deriving a new shared secret and identifier from the existing SA without repeating the full authentication procedure.

\textit{HashiCorp Vault} serves as a secure persistence layer for cryptographic material, rather than merely a key storage mechanism, and was selected for its encrypted persistent storage, fine-grained access control, and native auditing capabilities within a self-contained deployment model that is independent of the underlying infrastructure. Each node runs a dedicated Vault instance configured with file-based storage to ensure that cryptographic material persists across node restarts. The instance is bound exclusively to the local loopback interface, restricting access to processes running on the same node. Upon initialization, the Vault instance is configured with Shamir's Secret Sharing~\citep{shamir1979}, using a threshold of three out of five key shares. To support autonomous node initialization, the resulting unseal keys and root token are persisted with restricted file system permissions, and the instance remains sealed until explicitly unsealed by the internal components of the node. Secrets are managed using the Key-Value version 2 engine, which maintains a version history for all stored secrets. As illustrated in Fig.~\ref{fig:key_exchange}, the key material and its associated identifier are stored in the local Vault instance immediately after each key establishment, indexed by the peer address, and subsequently retrieved in response to identifier-based client requests without requiring a new key establishment. For post-quantum nodes, Vault additionally stores the IKE SA information established during the initial IKEv2 handshake, preserving the negotiated session state required to support efficient rekeying during subsequent key exchange operations. Access to the stored cryptographic material and session state is restricted to the local cryptographic components and HTTP receptor of each node, ensuring that sensitive cryptographic assets remain protected at rest and are accessible only to the internal components of the node through local authenticated interfaces.

\subsection{\textit{Quditto} Modeling Engine Package}

The Quditto modeling engine is the core component responsible for simulating quantum-level behavior of cryptographic key exchange systems and executing key generation protocols with realistic temporal characteristics. It comprises two main functional components: a message-broker handler that manages communication with Quditto nodes via RabbitMQ, and a quantum-behavior modeling core built on top of the NetSquid quantum simulator.

The engine operates as an asynchronous event-driven service, processing key generation requests for different QKD links in parallel while enforcing link-level serialization. This ensures that concurrent requests on different links proceed independently, improving overall throughput, yet maintains strict temporal ordering for multiple requests on the same link, accurately reflecting physical QKD system behavior where link resources cannot be simultaneously shared.

Upon receiving a "modeling request" message through the RabbitMQ broker, the engine validates the request by verifying that both nodes exist in the network and are neighbors on the same QKD link. If valid, the request is forwarded to the modeling core for protocol execution. By default, the engine features two QKD protocol implementations based on the BB84 scheme: \textit{BB84 with Eve} for eavesdropping scenarios, and \textit{Extended BB84} with user-configurable parameters to model realistic quantum hardware imperfections. Additional protocol implementations can be integrated as custom scripts based on NetSquid or any other quantum simulation platform, provided they produce cryptographic material and report realistic execution durations.

When processing a key request, the modeling engine executes the protocol implementation script repeatedly until sufficient cryptographic material is collected. Critically, the engine itself tracks how many bits remain to satisfy each request, eliminating the need for individual protocol implementations to accept key-length parameters. This separation of concerns simplifies protocol integration and allows developers to adopt flexible key generation strategies, such as maintaining internal buffers or computing bits on demand.

Once all required key material is gathered, the engine calculates and enforces the latency that would have been incurred by actual physical QKD hardware, based on protocol-specific parameters and link characteristics. The "modeling result" message, containing the generated key material and its unique identifier, is then published under the routing keys of both participating nodes after the calculated delay has elapsed.

Throughout this process, the modeling engine continuously records detailed logs capturing protocol interactions, timing events, and Quantum Bit Error Rate metrics. These logs are made available to users for post-execution analysis and debugging, enabling in-depth inspection of protocol behavior, performance measurement, and anomaly detection. In the cloud-native deployment, the modeling engine is instantiated as an independent Kubernetes pod with NetSquid and RabbitMQ installed by the orchestrator during initialization.

\subsection{\textit{Quditto} Orchestrator Package}
The \textit{Quditto} orchestrator package is responsible for the end-to-end deployment and configuration of the platform, from the provisioning of the underlying virtual infrastructure across cloud-hosted virtual machine and cloud-native Kubernetes environments to the initialization of all services comprising the network emulation environment. Provisioning performance has been further improved through the parallelization of Ansible playbook execution, substantially reducing the overall network instantiation and initialization time. The following subsections describe the resource provisioning capabilities and the end-to-end deployment workflow of the platform.

\subsubsection{Resource Provisioning and Execution Environment Preparation}
The orchestrator extends beyond the configuration of \textit{Quditto} components on pre-existing resources. In addition to deploying \textit{Quditto} on reachable physical or virtual machines, it provisions the virtual infrastructure required to host the platform in cloud-based environments. This functionality is built upon Terraform, an infrastructure-as-code tool used to define and provision virtual machines across different cloud platforms. The Terraform workflow is programmatically executed through a Python library using a declarative YAML-based description of the resources to be provisioned. This description specifies the required infrastructure parameters, including machine specifications, network interfaces, images, and access credentials, enabling automated virtual machine provisioning on cloud platforms such as OpenStack.

The provisioned resources support two execution models. In the first model, components are configured and executed directly on physical or virtual machines, following the original machine-based deployment model of \textit{Quditto}. In the second, the orchestrator introduces a cloud-native execution model in which components are deployed as containerized services on Kubernetes. To enable this execution model, the main \textit{Quditto} components have been packaged as Kubernetes artifacts through dedicated Helm Charts~\citep{helm}, which provide templated and versioned packages of Kubernetes resource definitions for the emulated node services and the modeling engine. These charts define the required Kubernetes resources and expose configurable parameters for component placement, service exposure, and container image selection. The container image parameter allows the orchestrator to instantiate components either from prebuilt images or from base images that are configured during the initialization of the network emulation environment.

To support the cloud-native deployment model, the orchestrator package incorporates the capability to bootstrap Kubernetes clusters on physical or virtual machines through KubeOne~\citep{kubeone}, a Kubernetes lifecycle management tool that automates cluster provisioning, upgrade, maintenance, and teardown while relying on \textit{kubeadm}~\citep{kubeadm}, the Kubernetes-native cluster bootstrapping utility, for the underlying cluster initialization and configuration. KubeOne operations are programmatically managed through Python-based logic, following the same integration approach used for Terraform-based virtual infrastructure provisioning. The package also includes functions to install the software dependencies required before cluster bootstrap, ensuring that this step is integrated into the orchestrated preparation of the execution environment rather than performed as a separate manual operation.

Both the Kubernetes bootstrap process and the \textit{Quditto} deployment follow a declarative YAML-based configuration approach. The cluster description provides the information required by KubeOne to bootstrap the cluster, including the machines comprising the cluster, their roles, and connectivity parameters. The \textit{Quditto} deployment description specifies the components to be deployed, their placement, configuration parameters, and exposed services. The orchestrator maps this description to the corresponding Helm chart values and deploys the appropriate Helm charts on the target cluster.

Additionally, the \textit{Quditto} deployment description includes placement parameters that allow components to be assigned to different Kubernetes clusters, enabling multi-cluster emulation scenarios. This execution model assumes only the availability of network connectivity between pods and services deployed across clusters, without imposing any specific multi-cluster networking solution. To provide this connectivity when required, the orchestrator package can dynamically install and configure multi-cluster networking plugins as part of the Kubernetes deployment workflow. The corresponding configuration parameters are specified in the YAML-based cluster description and mapped to the corresponding Helm chart values of the selected networking plugin. For instance, \textit{Submariner} can be used to provide transparent inter-cluster connectivity, allowing services running in different clusters to communicate without requiring modifications to the application-level logic of either the modeling engine or the emulated node services.

\subsubsection{Deployment Workflow of the \textit{Quditto} Network Emulation Environment}
\begin{figure*}
    \centering \includegraphics[page=1,width=\linewidth]{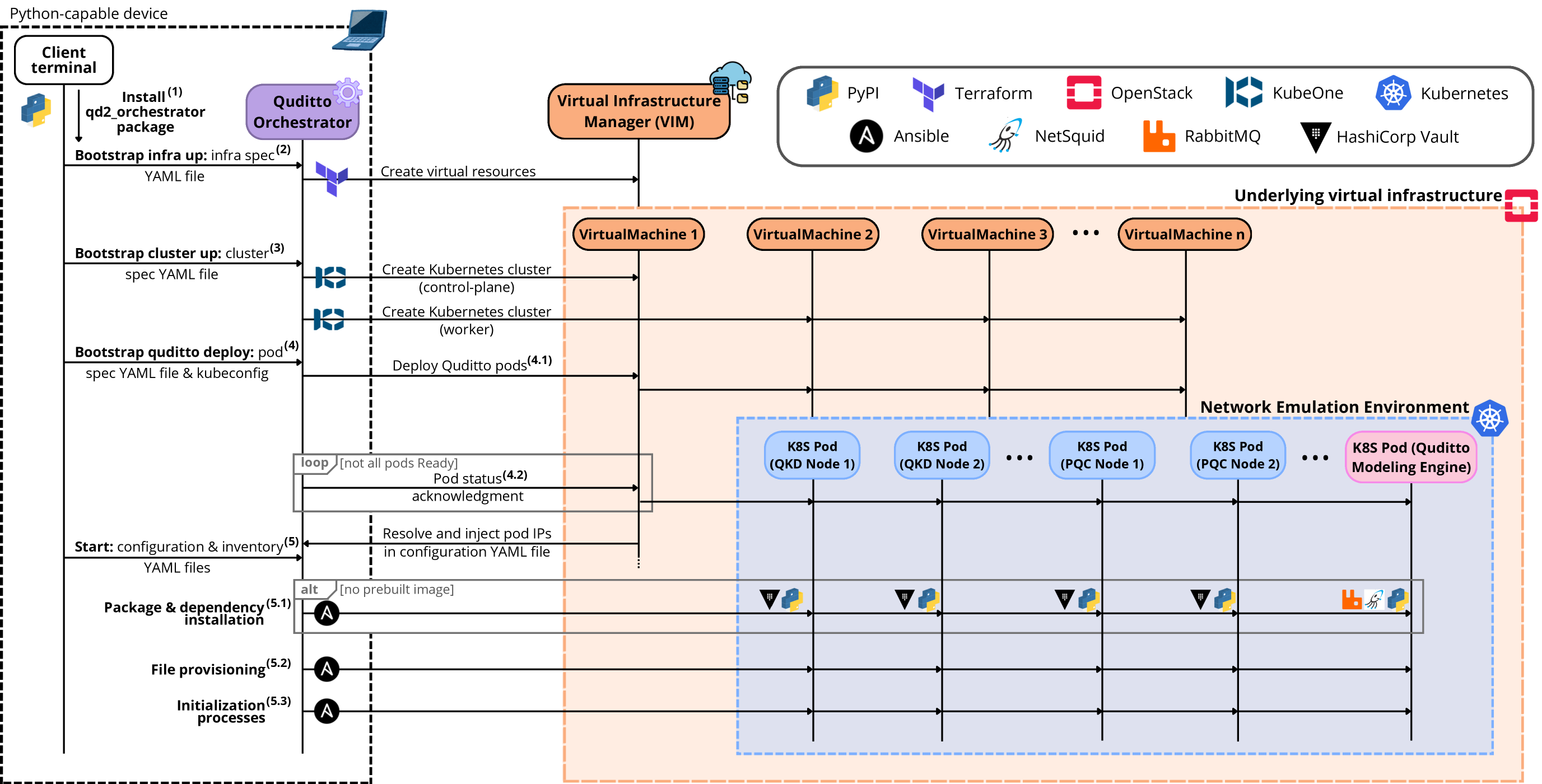}
    \caption{Automated deployment workflow of the \textit{Quditto} platform.}\label{fig:deployment}
\end{figure*}

Prior to the definition and instantiation of the emulated network, users may follow one of two workflows depending on the availability of existing infrastructure: either relying on pre-existing, user-managed machines with network connectivity or delegating the provisioning of the virtual infrastructure and the network emulation environment to the orchestrator. In the latter case, to fully exploit the advantages of the automated deployment process, the orchestrator implements parallelization across all deployment stages. The provisioning of the underlying virtual infrastructure, the deployment of \textit{Quditto} components as pods within Kubernetes clusters, and the subsequent dependency installation and service initialization procedures are all executed concurrently across the target nodes. Combined with the lightweight nature of containerized deployments, this parallelization strategy substantially reduces orchestration time and enables the platform to scale to networks comprising hundreds of nodes while maintaining acceptable deployment latency.
The complete automated deployment workflow is illustrated in Fig.~\ref{fig:deployment}.

Starting from a Python-capable client terminal, the \textit{Quditto} orchestrator package is installed via PyPI (stage 1). The orchestrator then provisions the required virtual machines for cloud environments such as OpenStack according to the corresponding infrastructure YAML specification (stage 2). Once the virtual machines are available, the orchestrator bootstraps a Kubernetes cluster across them, configuring the control-plane and worker nodes according to the cluster specification YAML file (stage 3). The provisioning process is configured through three declarative YAML files. The \texttt{vm-deployment} file specifies the parameters required to provision virtual machines, including machine specifications, network interfaces, and access credentials. The \texttt{k8s-deployment} file defines the configuration required to bootstrap the Kubernetes cluster, including node roles, network parameters, and cluster connectivity settings. The \texttt{qd2-deployment} file specifies the parameters required to deploy \textit{Quditto} pods within the cluster, including per-module Helm chart configuration and the Kubernetes NodePort services used to expose the ETSI GS QKD 014-compliant API port of each emulated network node on every cluster node, enabling external access to the key management interfaces. Notably, a virtual machine provisioned as part of the underlying virtual infrastructure may simultaneously host a \textit{Quditto} network node within the network emulation environment.

Once the cluster is operational, \textit{Quditto} pods are deployed within it (stage 4.1), including QKD nodes, PQC nodes, and the modeling engine, each instantiated as an independent Kubernetes pod forming part of the network emulation environment. Pod readiness is verified iteratively (stage 4.2). Once all pods are ready, their IP addresses are retrieved and injected into the configuration YAML file. \textit{Quditto} supports two deployment strategies for pod instantiation. The recommended approach uses prebuilt Docker images, one for each \textit{Quditto} component, available in the associated Docker Hub repository with all required dependencies preinstalled. Alternatively, when such images are unavailable, an automated dependency installation pipeline prepares the execution environment at deployment time, resulting in a fully operational system while preserving the same functional behavior under standard operating conditions. The deployment strategy is selected through the \texttt{version} field of the Helm chart entries in the \texttt{qd2-deployment} file.

Once the deployment strategy has been selected, the user defines the network to be emulated following the original \textit{Quditto} approach. The structure of the \texttt{inventory} file remains unchanged, providing the credentials used by Ansible to connect to the devices comprising the network emulation environment. The \texttt{configuration} file describing the network topology supports two types of nodes: quantum nodes, characterized by link lengths and the potential presence of eavesdroppers, and post-quantum nodes, defined by parameters such as the security level of the post-quantum digital signature algorithm.

During the emulated network initialization (stage 5), the orchestrator executes three substages via Ansible across all pods. When prebuilt Docker images are unavailable, the dependency installation pipeline is executed first (stage 5.1). The required Python dependencies are installed across all pods. Additionally, the \textit{HashiCorp Vault} client libraries are installed on node pods, whereas \textit{NetSquid} and \textit{RabbitMQ} are installed on the modeling engine pod. Once all dependencies have been installed, the orchestrator distributes the required Python scripts to each component (stage 5.2). These scripts implement the quantum or post-quantum key establishment logic according to the specified topology, together with an HTTP receptor for processing client requests and a Vault server instance for secure key management. Finally, the orchestrator starts all required services (stage 5.3), rendering the distributed system fully operational according to the user-provided network configuration. Once initialized, the emulated network enables clients to submit on-demand requests for quantum-safe cryptographic material through the ETSI GS QKD 014 specification.


\section{Validation} \label{sec:Validation}
To validate the capabilities of the proposed platform, this section presents a comprehensive evaluation structured along two complementary dimensions: orchestration performance and functional correctness. The evaluation of orchestration performance is motivated by the scalability constraints identified in the original \textit{Quditto} platform. In particular, as the size and complexity of the emulated networks increased, the time required to instantiate and configure the network components became a significant bottleneck, hindering the scalability of the platform for large-scale deployments. These limitations motivated the proposed orchestrator to reduce deployment overhead and enable the efficient management of larger hybrid network scenarios. Therefore, this evaluation focuses on characterizing the efficiency and scalability of the automated deployment workflow across networks of varying sizes, assessing its performance in terms of deployment time and  underlying computational infrastructure characteristics. The functional dimension assesses the ability of the platform to faithfully emulate the behavior of realistic hybrid quantum networks across diverse configurations and scenarios. All experiments were conducted on virtual machines hosted at the NEXTONIC laboratory~\citep{nextonic}, ensuring a controlled and reproducible execution environment.

\subsection{Orchestration Performance}
This evaluation investigates the orchestration performance of \textit{Quditto} as a function of network size, characterizing how the time required to deploy and initialize emulated hybrid networks scales with the number of nodes. The analysis identifies which stages of the orchestration process introduce overheads proportional to network growth and which remain invariant with respect to scale, delineating the practical scalability limits of the orchestration workflow. The results validate a core contribution of \textit{Quditto}: its capability to support the automated and scalable deployment of distributed hybrid network emulation environments without requiring pre-existing infrastructure.

For the experimental evaluation, the virtual infrastructure is automatically provisioned using the deployment workflow proposed in this work. The provisioning process operates on four pre-configured virtual machines deployed in a private OpenStack cloud environment. Although these virtual machines can also be provisioned by \textit{Quditto}, they are treated as part of the initial conditions of the experimental setup to isolate the evaluation of network-scale orchestration from the variability associated with environment-dependent infrastructure provisioning. By contrast, the Kubernetes cluster initialization constitutes a fixed one-time overhead that is independent of network scale and is therefore reported separately. Consequently, the provisioning process creates a minimal Kubernetes cluster configuration consisting of one control-plane node and two worker nodes, each deployed on a separate virtual machine. Across 20 repeated measurements, the cluster initialization time ranged from 162 s to 342 s, with a mean of approximately 230 s, reflecting the variability of Kubernetes cluster initialization in cloud environments. Once provisioned, this virtual infrastructure can support the deployment of multiple independent instances of the network emulation environment, thereby decoupling infrastructure provisioning from network scale. In this evaluation, each worker node hosts up to 100 Kubernetes pods, each emulating a quantum network node. Accordingly, network configurations of up to 100 nodes are deployed on a single worker node, whereas larger configurations are distributed across both worker nodes to maintain a balanced resource utilization.

\begin{figure}
    \centering
    \includegraphics[page=1,width=0.9\linewidth]{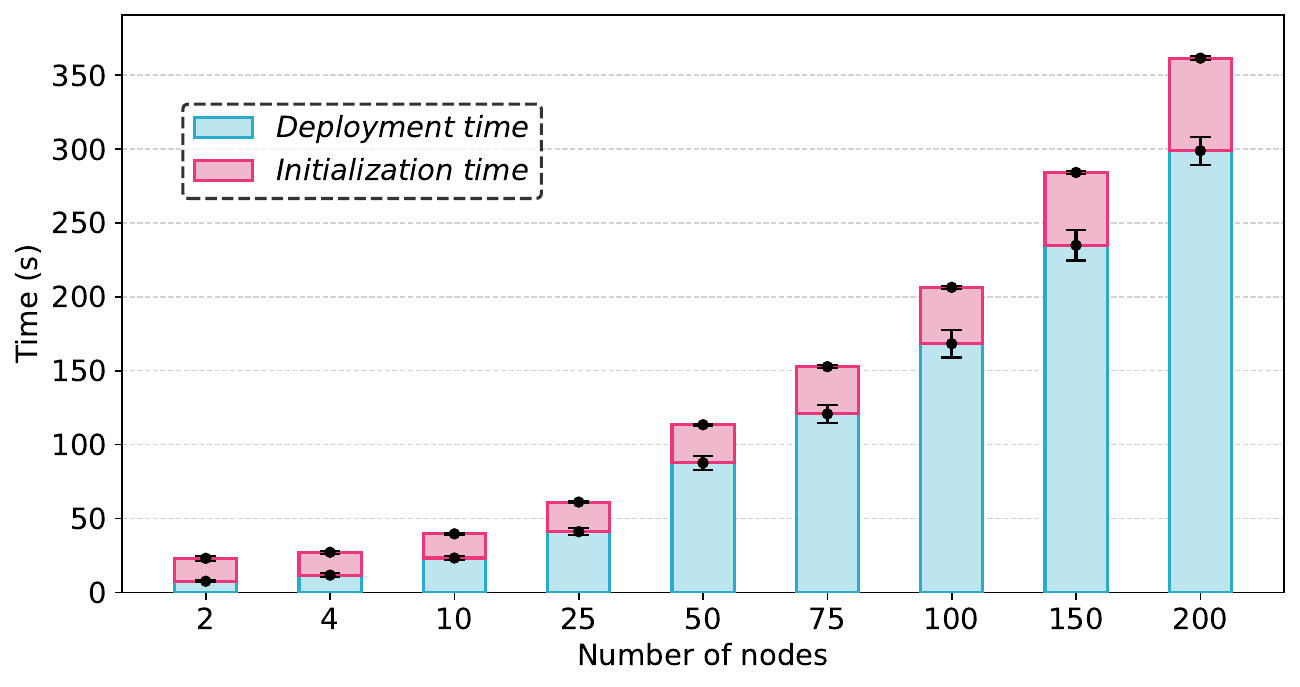}
    \caption{\textit{Quditto} orchestration time breakdown as a function of network size. Results correspond to 20 deployments per configuration, with error bars representing the 95\% confidence interval. Note that the horizontal axis is not linearly scaled.}\label{fig:scalability}
\end{figure}

All measurements reported in this section were obtained using the \textit{Quditto} prebuilt Docker images, which substantially reduce initialization overhead while ensuring consistent and reproducible execution across environments. This configuration also avoids the additional setup time associated with dependency installation, particularly for \textit{Vault}-related components. Consequently, this choice isolates orchestration performance from dependency management overhead, ensuring that the reported measurements accurately reflect the scalability of the system under the recommended deployment mode. The orchestration measurements were obtained using virtual machines with the following computational resources: the orchestrator node was allocated 48 vCPUs, 80 GB of RAM, and 50 GB of storage, while each cluster node was provisioned with 24 vCPUs, 32 GB of RAM, and 50 GB of storage. Fig. \ref{fig:scalability} presents the time breakdown of the orchestration process as a function of the number of emulated quantum network nodes, decomposed into its two variable-duration stages.

The overall orchestration time increases with network size, confirming that the orchestration workload primarily scales with node-related operations. The network emulation environment deployment stage constitutes the dominant contributor to total orchestration time, particularly for larger network configurations. The observed scaling behavior is consistent with sublinear growth, with the total deployment time increasing from 7.6 s for a 2-node network to 298.8 s for a 200-node network, corresponding to a 39-fold increase in deployment time for a 100-fold increase in network size. This sublinear trend reflects the high degree of parallelization achieved through concurrent pod instantiation. However, the growth rate increases progressively for larger configurations, as Kubernetes scheduling latency and resource allocation pressure accumulate with the number of managed pods, despite Ansible executing operations concurrently in batches. In contrast, the node initialization stage exhibits a markedly lower growth rate, increasing from 15.5 s for a 2-node network to 62.6 s for a 200-node network, consistent with the parallelization of per-node provisioning and service startup, which distributes the initialization workload across nodes concurrently. Notably, even at the largest evaluated scale of 200 nodes, the total orchestration time remains within approximately six minutes, demonstrating that \textit{Quditto} supports the automated deployment of large-scale emulated networks within practical operational times.

These results demonstrate a significant improvement in deployment efficiency compared to the original \textit{Quditto} platform. Although an absolute comparison with the deployment times of the original platform is not possible due to the absence of documented infrastructure specifications for the previous setup, a meaningful qualitative assessment can still be drawn. As shown in Fig.~3 of the original \textit{Quditto} work, node initialization did not scale efficiently with the number of nodes, effectively limiting practical deployments to networks of up to 20 nodes. This limitation has been addressed through two targeted optimizations: the adoption of prebuilt Docker images, which eliminates sequential per-node dependency installation, and the parallelization of the network emulation environment deployment and node initialization stages. Together, these improvements extended the practical deployment scale of the platform from small configurations of up to 20 nodes to large-scale deployments of 200 or more nodes within approximately six minutes, representing an order-of-magnitude increase in the supported network scale.

To assess the robustness of the orchestration workflow under resource-constrained conditions, a complementary set of experiments was conducted on machines with significantly reduced computational resources. In this configuration, the orchestrator was allocated 2 vCPUs, 4 GB of RAM, and 50 GB of storage, while each cluster node was configured with 4 vCPUs, 8 GB of RAM, and 50 GB of storage. For an intermediate network configuration of 50 nodes, the reduced-resource setup required approximately 96 s for network emulation environment deployment and 65 s for node initialization. In comparison, under the high-resource configuration shown in Fig. \ref{fig:scalability}, the same stages required approximately 87 s and 25 s, respectively. Notably, the impact of reduced computational resources is asymmetric across the two stages: the network emulation environment deployment time increases by approximately 10\% (from 87 s to 96 s), whereas node initialization time increases by a factor of 2.6 (from 25 s to 65 s). This disparity suggests that the deployment stage is primarily constrained by Kubernetes scheduling overhead rather than by available computational resources, while the initialization stage, which involves dependency installation and service initialization, is more sensitive to CPU and memory availability. These results indicate that large-scale emulated hybrid network configurations can be deployed within acceptable time frames on modest infrastructure, thereby demonstrating the feasibility of operating the platform without requiring dedicated high-performance computing resources.

\subsection{Functional Validation of Hybrid Networks}

This validation evaluates the functional performance of a key establishment protocol executed between a post-quantum node and a quantum-enabled node through a hybrid network under realistic operating conditions. To this end, a hybrid QKD-PQC network is deployed following a spine-leaf topology (described in detail below), over which a key establishment is performed between an access-level post-quantum site and a spine-level quantum site while the network is subjected to background traffic, i.e., concurrent request load emulating realistic network usage. The study characterizes the impact of this background traffic on the execution of the protocol by systematically measuring the network load alongside the temporal performance of the primary key establishment process. Timestamps are collected for each relevant event, including master key generation, \texttt{get\_key} and \texttt{get\_key\_with\_ID}, master key encryption and decryption, and completion of the establishment, enabling a detailed breakdown of the whole process. The analysis identifies how concurrent traffic influences protocol efficiency, thereby establishing the robustness and timing reliability of the establishment mechanism in hybrid environments.

\begin{figure}
    \centering
    \includegraphics[page=1,width=0.8 \linewidth]{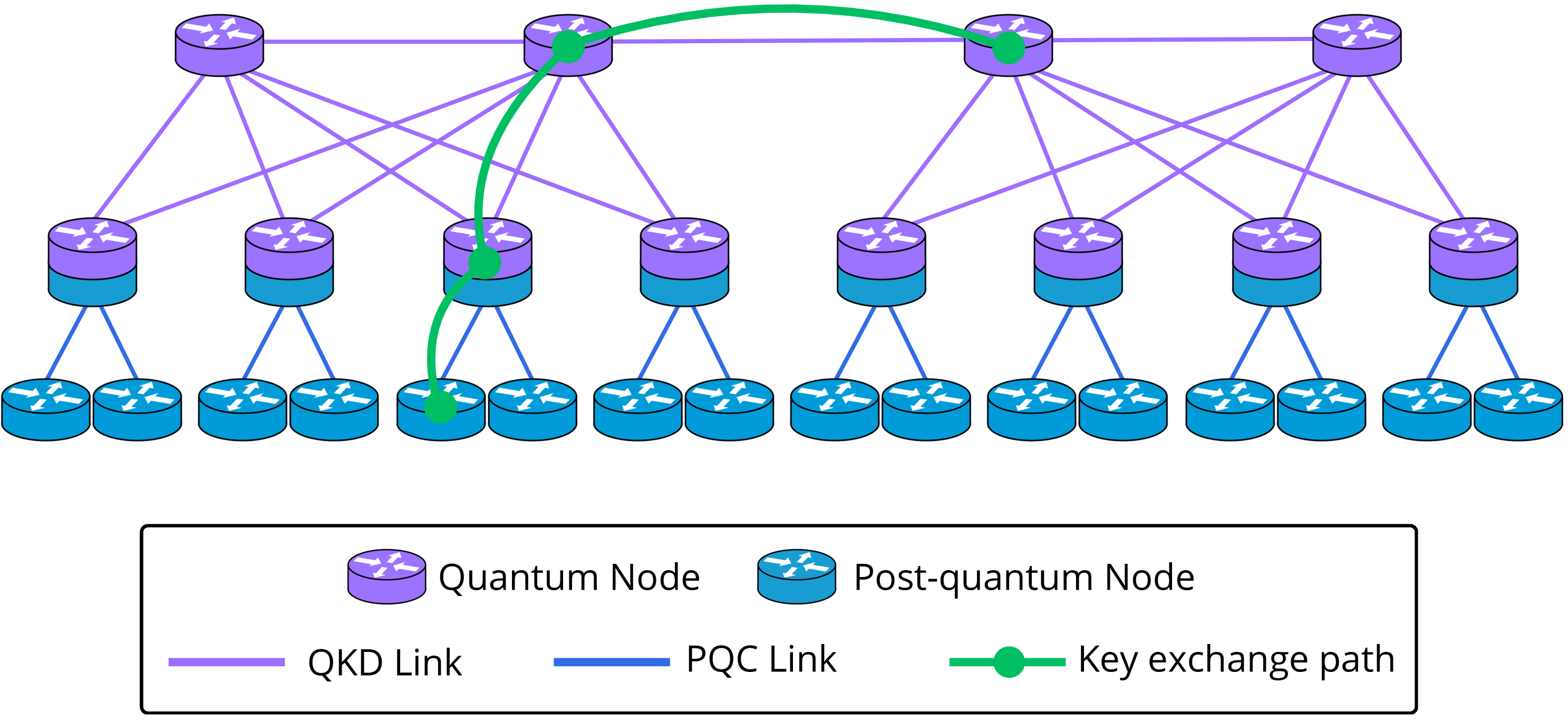}
    \caption{Schematic of the hybrid QKD-PQC spine-leaf network deployment used for the functionality test. The network comprises 36 nodes and 35 communication links, including 19 QKD links and 16 PQC links. The nodes are distributed across three logical levels: the spine level, with 4 QKD nodes; the leaf level, with 8 QKD and 8 PQC nodes; and the access level, with 16 PQC nodes. The path followed by the key establishment protocol during the functionality test is highlighted in green.}\label{fig:Network}
\end{figure}

However, prior to presenting the measurement results, the network deployment used in this validation is described. The chosen scenario is a hybrid network with a spine-leaf topology architecture. This specific architecture was selected to evaluate the performance of an evenly distributed hybrid environment with both QKD and PQC loads. As can be seen in Fig. \ref{fig:Network}, although there are more post-quantum sites per se than quantum sites, the number of quantum and post-quantum links is nearly the same, slightly favoring quantum links. A real-world justification for this deployment can be drawn from large-scale, security-critical infrastructures, where highly secure and physically protected core links employ QKD, intermediate aggregation nodes use hybrid QKD-PQC to ensure interoperability and resilience, and edge or access nodes rely on PQC to enable scalable and cost-effective connectivity.

\begin{figure}
    \centering
    \includegraphics[page=1,width=0.9\linewidth]{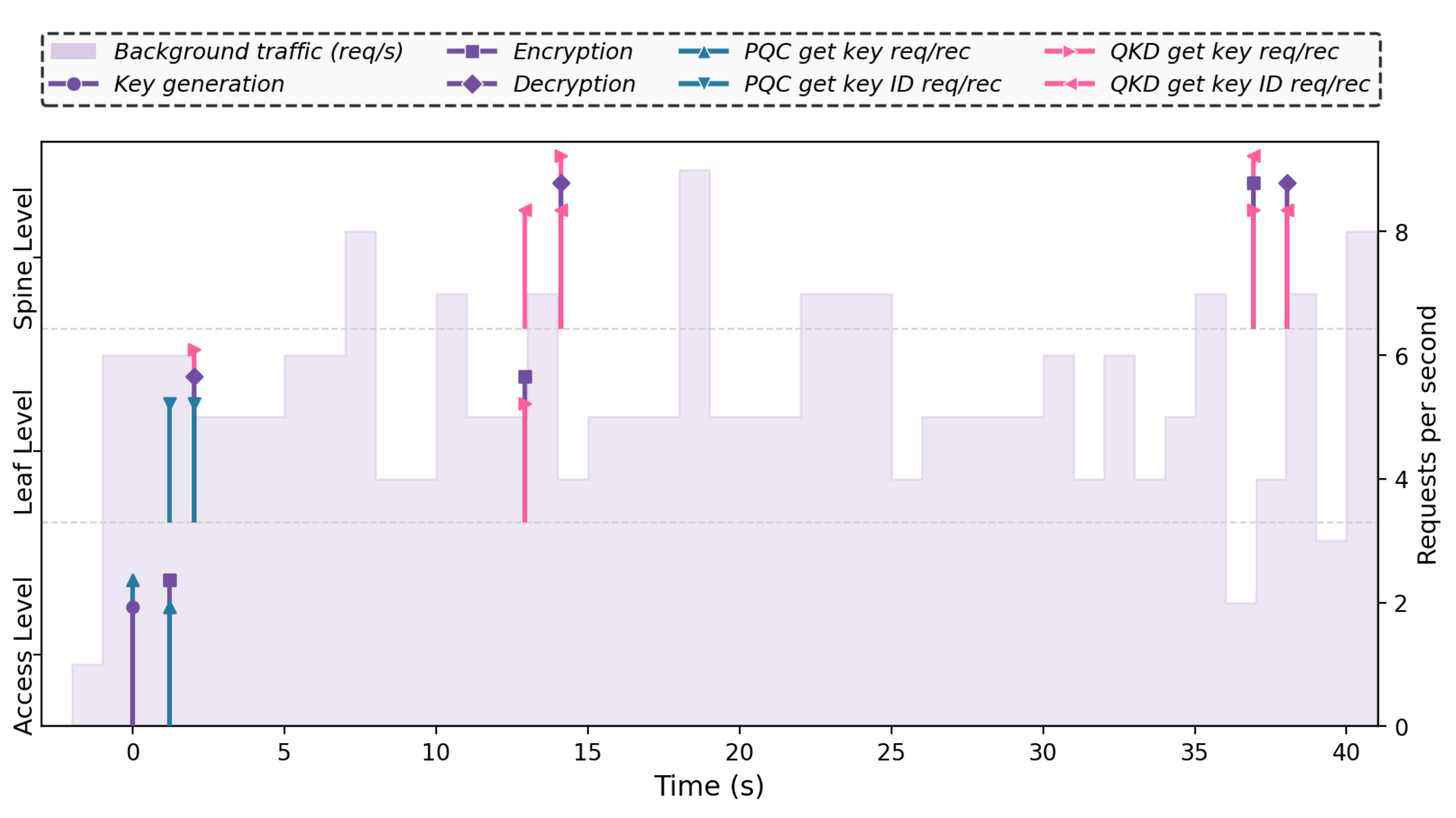}
    \caption{Timeline of the establishment of a 256-bit key between a post-quantum site and a quantum site in an emulated hybrid network. Vertical markers indicate key events (deltas) at the Access (PQC), Leaf (hybrid), and Spine (QKD) levels, including key generation, key request, encryption, and decryption operations. The shaded area represents the background network traffic intensity, measured in requests per second (req/s), throughout the protocol execution.}\label{fig:functionality}
\end{figure}

The measurements shown in Fig.~\ref{fig:functionality} present the timestamps corresponding to a secure key establishment protocol between two nodes belonging to different technological planes, i.e., between a post-quantum and a quantum cryptographic node. The figure shows the creation of the master key for the exchange as the initial timestamp, as well as the encryption and decryption of this key at each node hop, together with the timestamps corresponding to the initialization and resolution of the point-to-point key establishments between neighboring sites used to generate the encryption keys that ensure secure distribution. Additionally, the figure includes the background traffic of the emulated network, measured in requests per second (shaded purple area, right axis), and is divided into three horizontal sections, each corresponding respectively to the three levels of the spine-leaf architecture shown in Fig.~\ref{fig:Network}, namely the \emph{Access}, \emph{Leaf}, and \emph{Spine} levels, plotted bottom-to-top.

At the \emph{Access Level}, the protocol starts at $t = 0.000\,\text{s}$ with the generation of the master key, marked by a purple circle ($\bullet$). The PQC key-request round-trip is shown by the blue upward-pointing triangles ($\blacktriangle$): the request is sent at $t = 0.000\,\text{s}$ and the response is received at $t = 1.210\,\text{s}$. The resulting AES encryption is marked by a purple square ($\blacksquare$) at $t = 1.211\,\text{s}$. Moving to the \emph{Leaf Level}, the PQC key-ID request round-trip is shown by the blue downward-pointing triangles ($\blacktriangledown$), sent at $t = 1.211\,\text{s}$ and received at $t = 2.020\,\text{s}$, followed immediately by the PQC decryption (purple diamond, \rotatebox[origin=c]{45}{$\blacksquare$}) at $t = 2.020\,\text{s}$. The first QKD key-request round-trip, marked by pink right-pointing triangles ($\blacktriangleright$), is then sent at $t = 2.020\,\text{s}$ and received at $t = 12.915\,\text{s}$, after which the data is re-encrypted with the QKD key (purple square) at $t = 12.916\,\text{s}$. At the \emph{Spine Level}, the first QKD key-ID request round-trip is marked by pink left-pointing triangles ($\blacktriangleleft$), sent at $t = 12.916\,\text{s}$ and received at $t = 14.103\,\text{s}$; the QKD decryption of the first hop (purple diamond) and the second QKD key request (pink right-pointing triangle, sent) occur within the same millisecond, at $t = 14.103\,\text{s}$, and are shown stacked front-to-back in chronological order to keep them distinguishable. The second QKD key-request round-trip completes when the response is received at $t = 36.941\,\text{s}$ (pink right-pointing triangle), followed by the second QKD encryption (purple square) at $t = 36.942\,\text{s}$. Finally, the second QKD key-ID request round-trip, marked by pink left-pointing triangles, is sent at $t = 36.942\,\text{s}$ and received at $t = 38.032\,\text{s}$, and the protocol concludes with the final decryption (purple diamond) at $t = 38.033\,\text{s}$.

The results of the measurement show that the average background traffic during the main establishment was 325 requests per minute. Taking into account this request load, the main key establishment took exactly $\Delta t = 38.033\,\text{s}$ to complete its secure distribution from the post-quantum access node to the quantum spine node. These measurements demonstrate the viability of establishing cryptographic keying material between nodes in different logical planes of the hybrid emulated network. 

These results, as well as the ones obtained in the scalability performance test shown in Fig. \ref{fig:scalability}, were measured on virtual machines with considerable computational resources. For the purpose of highlighting results of particular relevance, it can be noted that in the PQC key establishment, both the generation of the encryption key and its subsequent recovery, are fulfilled in $\Delta t = 1.210\,\text{s}$ and $\Delta t = 0.820\,\text{s}$, respectively. However, even though the recovery of the quantum key after its generation is fulfilled in about the same amount of time ($\Delta t_1 = 1.187\,\text{s}$, $\Delta t_2 = 1.090\,\text{s}$), the generation of said key is by far the longest process in the whole establishment: the first generation with $\Delta t = 10.895\,\text{s}$ and the last generation with $\Delta t = 22.838\,\text{s}$.  

Both the disparity of length between the post-quantum and quantum key generations as well as the difference of time between both quantum generations in the process can be explained by the same phenomenon: generating quantum cryptographic material of a specific length is highly variable, as it depends on pure probability in the quantum measurements. It is also noteworthy that for the sake of observing the full performance of the network under a considerable workload, the quantum cryptographic material used to generate the QKD encryption keys was generated on demand rather than pre-computed, in order to observe the full operational behavior of the network under a representative workload.

\section{Conclusion} \label{sec:Conclusions}
This work presented \textit{Quditto} as an emulation platform that addresses the automation, scalability, and hybridization challenges identified in the original platform, thereby filling a significant gap in the state of the art regarding reproducible emulation of hybrid quantum-safe network infrastructures. Four principal contributions were introduced: a cloud-native orchestrator supporting fully automated virtual infrastructure provisioning across cloud and multi-cluster Kubernetes environments; a parallelized deployment workflow enabling networks of up to 200 nodes to be deployed within a practically acceptable time frame; the integration of post-quantum nodes implementing IKEv2-based key establishment with ML-KEM and ML-DSA, enabling native emulation of hybrid QKD-PQC topologies; and a secure key management module built on \textit{HashiCorp Vault}, providing auditable, integrity-protected, and persistent storage of cryptographic material.

The experimental validation substantiated these contributions along two complementary dimensions. The orchestration performance evaluation demonstrated sublinear scaling of deployment time with network size, extending the practical deployment scale of the platform from the small-scale configurations supported by the original \textit{Quditto}, limited to approximately 20 nodes, to large-scale deployments exceeding 200 nodes in approximately six minutes. Furthermore, the robustness assessment under resource-constrained conditions showed that this scalability does not require high-performance computing infrastructure, with the deployment stage remaining primarily constrained by Kubernetes scheduling overhead rather than by computational resource availability. The functional validation complemented these results by confirming the correct end-to-end operation of a hybrid key exchange over a representative spine-leaf topology, demonstrating reliable interoperability between the quantum and post-quantum planes under realistic background traffic conditions,
 and characterizing the distinct temporal behavior of QKD- and PQC-based key generation within a unified key establishment procedure. Collectively, these results confirm that the proposed platform closes a significant gap identified in the literature: the absence of emulation and orchestration frameworks capable of reproducing hybrid QKD-PQC network behavior at scale without requiring dedicated physical quantum infrastructure or manually provisioned virtual resources. By coupling automated infrastructure provisioning with a standards-based hybrid key establishment framework, \textit{Quditto} provides the research community with an accessible, reproducible, and extensible testbed for evaluating quantum-safe cryptographic architectures under realistic and controllable conditions.

 Future work will explore several directions building on this foundation. First, the evaluation of additional network topologies and larger-scale hybrid deployments would further characterize the interplay between orchestration overhead and the emulated network structure. Second, the integration of adaptive operational modes capable of switching dynamically between QKD-only, PQC-only, and hybrid configurations, as explored in recent proposals surveyed in Section~\ref{sec:hybrid}, would broaden the applicability of the platform to heterogeneous and evolving network conditions. Finally, progressive integration with physical QKD devices, as envisioned in the architectural design of the platform, would enable systematic validation of emulated results against real-world quantum hardware, further bridging the gap between controlled experimentation and practical deployment.









\section*{Acknowledgements}
The present work has been supported by the 6G-INSPIRE project, PID2022-137329OB-C42, funded by MCIN/AEI/ 10.13039/501100011033/ and by the EU Horizon Europe project Quantum Security Networks Partnership (QSNP), under grant 101114043.

\bibliographystyle{cas-model2-names}

\bibliography{cas-refs}



\end{document}